\documentclass{aa}  
\usepackage{graphicx}
\usepackage{txfonts}
\usepackage{afterpage}
\begin{document}

   \title{WSRT Faraday tomography of the Galactic ISM at $\lambda \sim 0.86$ m}

   \subtitle{I. The GEMINI data set at $(l,b)$=(181\degr,20\degr)}

   \author{D.H.F.M. Schnitzeler\inst{1,2}
          \and
          P. Katgert\inst{1}
          \and
          A.G de Bruyn\inst{3,4}
          }

   \offprints{D. Schnitzeler}

   \institute{Leiden Observatory, Leiden University, P.O. Box 9513, 2300 RA Leiden, The Netherlands
         \and
             Australia Telescope National Facility, CSIRO, Marsfield, NSW 2122, Australia\\
              \email{dominic.schnitzeler@csiro.au}
         \and
             ASTRON, P.O. Box 2, 7990 AA Dwingeloo, The Netherlands
         \and 
             Kapteyn Institute, P.O. Box 800, 9700 AV Groningen, The Netherlands
             }

   \date{}

  \abstract
   {}
  {We investigate the properties of the Galactic ISM by applying
   Faraday tomography to a radio polarization data set in the direction
   of the Galactic anti-centre.}
  {We address the problem of missing large-scale structure in our data,
   and show that this does not play an important role for the results we present.}
  {The main peak of the Faraday depth spectra in our
  data set is not measurably resolved for about 8\% of the
  lines of sight. An unresolved peak indicates a separation between
  the regions with Faraday rotation and synchrotron emission. However,
  cosmic rays pervade the ISM, and synchrotron emission would
  therefore also be produced where there is Faraday rotation. We
  suggest that the orientation of the magnetic field can separate the
  two effects. By modelling the thermal electron contribution to the
  Faraday depth, we map the strength of the magnetic field component
  along the line of sight. Polarized point sources in our data set have
  rotation measures that are comparable to the Faraday depths of the
  diffuse emission in our data. Our Faraday depth maps show narrow
  canals of low polarized intensity. We conclude that depolarization
  over the telescope beam produces at least some of these
  canals. Finally, we investigate the properties of one conspicuous
  region in this data set and argue that it is created by a decrease in
  line-of-sight depolarization compared to its surroundings.  
} 
   {}


   \maketitle
%

\section{Introduction}
Faraday rotation, the rotation of the plane of linear polarization due
to the birefringence of a magneto-ionic medium, provides a powerful
tool for exploring the magnetic universe, complementing effects such
as Zeeman splitting and the alignment of ellipsoidal dust grains
perpendicular to magnetic field lines. From external galaxies we have
learnt about the global structure of the magnetic field, about its
alignment with the galactic spiral arms, and of the importance of the
magnetic pressure in ISM dynamics, which is comparable to the thermal pressure
in the cold and warm neutral phases of the ISM (see Beck
\cite{beck07a}, \cite{beck07b}).


In the Milky Way, rotation measures of pulsars, combined with
dispersion measures, yield a rather complicated picture of the
large-scale Galactic magnetic field, with evidence for several field
reversals (see e.g. Han et al. \cite{han06}). Rotation measures of
extragalactic sources from the International Galactic Plane Survey
(IGPS) indicate a simpler structure of the large-scale field, without
much evidence for many field reversals (Brown et al. \cite{brown07}).

Studies of Faraday rotation of the diffuse Galactic ISM offer a big advantage over other investigations based on pulsars and extragalactic sources, which is that the diffuse emission is visible in all directions. This has been spectacularly demonstrated by the 1.41 GHz survey carried out with the DRAO 26 m telescope by Wolleben et al. (\cite{wolleben06}). 
A milestone in this type of work is the polarization survey by Brouw \& Spoelstra in the late 1970s (\cite{brouwspoelstra76} and Spoelstra \cite{spoelstra84}), which has the largest sky coverage in rotation measures of the diffuse emission to date. 
Since the 1990s interferometric studies at low frequencies have led to new insights in e.g. the properties of the ISM on small scales, and on turbulence in the Galactic ISM (Wieringa et al. \cite{wieringa93}; Haverkorn et al. \cite{haverkorn04}, \cite{haverkorn06}).

Recently, a method called Faraday tomography has been developed with which one can study the relative distribution of synchrotron-emitting and Faraday-rotating regions along the line of sight (see Brentjens \& De Bruyn \cite{brentjensbruyn05}). Application of this technique for studying the Galactic ISM is a very new field of research, and not much literature is available on this subject.
In Schnitzeler et al. (\cite{dominic07b}) we discussed a low-frequency data set that covers $\approx\ 50^{\Box}$ in the direction of ($l,b$) = (181\degr, 20\degr), that we obtained with the WSRT. We concluded that many lines of sight contain at least one Faraday screen, a region in which Faraday rotation and synchrotron emission are not mixed. This result appears counterintuitive, since one would expect synchrotron-emitting cosmic rays to be present everywhere there are magnetic fields. 
In this paper we present the full analysis of this data set.

Faraday tomography promises to become an interesting and exciting way to investigate the Galactic ISM that will become more and more available in the near future, with the advent of new surveys and radio telescopes that combine high spatial resolution with high frequency resolution, like the GALFACTS survey that is carried out with the Arecibo telescope (Taylor \cite{taylor04}), the LOFAR array (R\"ottgering et al. \cite{rottgering06}), and the MWA SKA-precursor (Bowman et al. \cite{bowman06}). A better understanding of the Galactic foreground will also be useful for studies of the cosmic microwave background.

This paper is the first in a series of articles in which we discuss different regions in the second Galactic quadrant. Here we discuss some of the observational features that we encounter, and we extend some of the techniques that have been used to study the diffuse Galactic ISM with a relatively small number of frequency channels to the many-channel regime.
First, we give a short overview of Faraday tomography in Sect. \ref{ch3_Sect: Faraday tomography}, and we illustrate its potential with a couple of examples. In Sect. \ref{ch3_Sect: data_characteristics} we discuss the observational characteristics of the GEMINI data set, and in Sect. \ref{ch3_Sect: analysis} we present the GEMINI data. We show that Faraday modulation in the foreground can convert polarized emission on large angular scales to smaller angular scales that can be picked up by an interferometer that has no short baselines. We derive the magnetic field strengths implied by the observed Faraday depths of the diffuse emission in Sect. \ref{ch3_Sect: magnetic field}, and in Sect. \ref{ch3_Sect: point sources} we compare the Faraday depths we derive for the GEMINI data set with the rotation measures that we find for polarized point sources. One conspicuous feature in the GEMINI data are the dark canals in the maps at constant Faraday depth, and we investigate their origin in Sect. \ref{ch3_Sect: canals}. In Sect. \ref{ch3_Sect: blob} we present an explanation for a large and bright polarized region in our data set.

\section{Notes on Faraday tomography}\label{ch3_Sect: Faraday tomography}
\subsection{Faraday tomography}\label{ch3_Sect: intro_FT}
The wide coverage in $\lambda^2$ space of our data (from
0.6 m$^2\lesssim\ \lambda^2\ \lesssim\ 0.9$ m$^2$), combined with the
relatively small channelwidth ($\delta\lambda^2\ \approx\ 1.5\times10^{-3}$ m$^2$), allows us to do Faraday tomography,
also known as Rotation Measure Synthesis (see e.g. Brentjens \& De
Bruyn \cite{brentjensbruyn05}). In this section we introduce the
concept of Faraday tomography, and we illustrate how it can be used to
study the properties and the relative distribution of regions with 
Faraday rotation and synchrotron emission along the line of sight.

The observed Stokes parameters $Q(\lambda)$ and $U(\lambda)$ are the 2 orthogonal components of the polarization vector $\vec{P}(\lambda)$ = $Q(\lambda)+\mathrm{i}U(\lambda)$. $\vec{P}(\lambda)$ is the vector sum of all polarization vectors that are emitted along the line of sight and that are Faraday rotated in the ISM between the point of emission and the observer:

\begin{eqnarray}
\vec{P}(\lambda)\ =\ \int\limits_{\mathrm{0}}^{\infty} \vec{P}(\mathrm{x})\ \mathrm{e}^{2\mathrm{i}\mathcal{R}\mathrm{(}\mathrm{x)}\lambda^2}\mbox{dx}\ =\ \int\limits_{-\infty}^{\infty} \vec{P}(\mathcal{R})\ \mathrm{e}^{2\mathrm{i}\mathcal{R}\lambda^2}\mbox{d}\mathcal{R}
\label{ch3_p_lambda2.eqn}
\end{eqnarray}

\noindent
where the first integral is over physical distance `x', and the Faraday depth of point `x'

\begin{eqnarray}
\mathcal{R}(x)\ [\mathrm{rad/m}^2]\ =\ 0.81\int_{\mathrm{source\ at\ `x'}}^{\mathrm{observer}} n_e\ [\mathrm{cm}^{-3}]\ \vec{B}\ [\mu\mathrm{G}]\cdot\mbox{d}\vec{l}\ [\mathrm{pc}]
\label{ch3_curlyR.eqn}
\end{eqnarray}

\noindent
measures the total amount of Faraday rotation between the point of emission `x' and the observer.
In the second integral of Eqn. \ref{ch3_p_lambda2.eqn} we have replaced the integral over physical distance by an integral over Faraday depth. If there is a one-to-one correspondence between `x' and $\mathcal{R}$, this change in coordinate is trivial. However, a change in the direction of the magnetic field component along the line of sight $B_{\|}\ =\ \vec{B}\cdot \mbox{d}\vec{l}$ assigns the same $\mathcal{R}$ to different `x'. To go back from $\mathcal{R}$ to `x' thus requires assumptions on the magnetic field geometry. In a future article we will return to this issue. Faraday tomography is based on the inversion of the second integral in Eqn. \ref{ch3_p_lambda2.eqn}: 




\begin{eqnarray}
\vec{P}(\mathcal{R}) & = & K\int\limits_{-\infty}^{\infty} W(\lambda^2)\ \vec{P}(\lambda^2)\ \mathrm{e}^{-2\mathrm{i}\mathcal{R}\lambda^2}\mbox{d}\lambda^2 \label{ch3_p_r.eqn} \\
\mathrm{where} & & \nonumber \\
K & = & \left(\int\limits_{-\infty}^{\infty} W(\lambda^2)\ \mbox{d}\lambda^2\right)^{-1} \label{ch3_norm.eqn}
\end{eqnarray}

\noindent
produces the correct normalization for $\vec{P}(\mathcal{R})$.
$W(\lambda^2)$ = 1 if the $\lambda^2$ has been observed, otherwise
$W(\lambda^2)$ = 0.  This inversion can be interpreted either as a
`coherent addition' of the observed $\vec{P}(\lambda)$ by using an
assumed Faraday depth to derotate the polarization vectors, or,
similarly, as a Fourier transform from $\lambda^2$ space to
$\mathcal{R}$ space. Note that the integral is also calculated over
negative values of $\lambda^2$, which is physically impossible. If the
measured $P(\lambda^2)$\ are in K (or mJy/beam), then also the $P(\mathcal{R})$\ will be in K
(or mJy/beam), or, more precisely, in K/RMSF width (or mJy/beam/RMSF
width). If we want to compare different data sets, with different
RMSFs, it is necessary to convert this quantity to K/rad/m$^2$ instead
of K/RMSF (equivalently for mJy/rad/m$^2$). Here we only consider one
data set, and we did not convert the measured $P(\mathcal{R})$\ to K/rad/m$^2$.


The complex polarization vector $\vec{P}(\lambda)$ can be written in
terms of the observed polarized intensity $P(\lambda)$ and
polarization angle $\Phi(\lambda)$ as $\vec{P}(\lambda)\ =\ P(\lambda)\
\mathrm{e}^{2\mathrm{i}\Phi(\lambda)}$. Similarly,
$\vec{P}(\mathcal{R})\ =\ P(\mathcal{R})\
\mathrm{e}^{2\mathrm{i}\Phi(\mathcal{R})}$, where $P(\mathcal{R})$ is
the intensity of the polarized emission at Faraday depth
$\mathcal{R}$, and $\Phi(\mathcal{R})$ is the orientation of the
electric field vector of the synchrotron radiation emitted at Faraday
depth $\mathcal{R}$, which is perpendicular to the local direction of the
magnetic field at that Faraday depth. A Faraday depth spectrum (or
$\mathcal{R}$ spectrum) can be constructed by calculating
$P(\mathcal{R})$ and $\Phi(\mathcal{R})$ for many $\mathcal{R}$ values.

The Fourier-transform nature of Eqn. \ref{ch3_p_r.eqn} implies that it shares some characteristics with other Fourier-transform based methods like radio interferometry. Eqns. 61 to 63 from Brentjens \& De Bruyn (\cite{brentjensbruyn05}) describe this behaviour quantitatively, and for convenience we reproduce these equations here. We assume here that the observations uniformly span the $\lambda^2$ range from $\lambda^2_{\mathrm{min}}$ to $\lambda^2_{\mathrm{max}}\ =\ \lambda^2_{\mathrm{min}}+\Delta\lambda^2$ with weight 1, and with weight 0 outside this range.

The finite extent in the $\Delta\lambda^2$ coverage of the data introduces an instrumental response along the $\mathcal{R}$\ axis, known as the RMSF (Rotation Measure Spread Function), with which the P($\mathcal{R}$) spectrum is convolved. The shape of the (normalised) RMSF is given by

\begin{eqnarray}
\mathrm{RMSF}\ (\mathcal{R})\ =\ K \int\limits_{-\infty}^{\infty} W(\lambda^2)\ \mathrm{e}^{-2\mathrm{i}\mathcal{R}\lambda^2}\mbox{d}\lambda^2
\label{ch3_rmsf.eqn}
\end{eqnarray}

\noindent
. The FWHM of the RMSF depends on the total range in $\lambda^2$ space covered by the observations, $\Delta\lambda^2$, according to

\begin{eqnarray}
\mathrm{FWHM\ [radians/m^2]}\ =\ \frac{3.8}{\Delta\lambda^2}
\label{ch3_fwhm.eqn}
\end{eqnarray}

\noindent
. We replaced the factor of $2\sqrt{3}$ from the original paper by Brentjens \& De Bruyn by 3.8, because 3.8 is closer to the FWHM of the sinc RMSF response than $2\sqrt{3}\ \approx\ 3.5$.

\begin{figure}[t]
\resizebox{\hsize}{!}{\includegraphics[width=8.5cm]{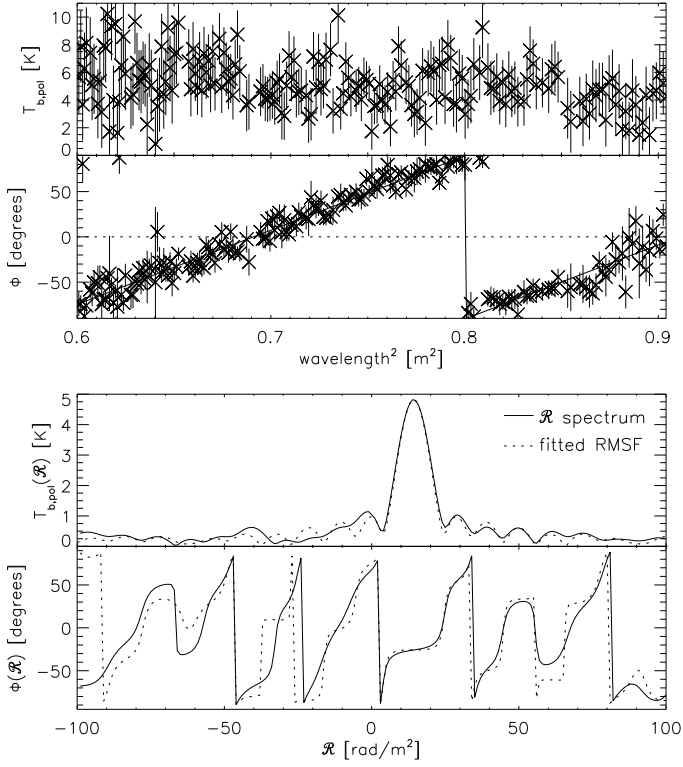}}
\caption{P$(\lambda^2)$ and $\Phi(\lambda^2)$ distributions (top 2
  panels), and P($\mathcal{R}$) and $\Phi(\mathcal{R})$ spectra
  (bottom 2 panels) for a line of sight with a barely resolved main
  peak that is similar to a Faraday screen.  In the text we describe
  how we derotated the polarization angles in the bottom panel of this
  figure.  The $\chi^2_{\mathrm{red}}$ of the RM fit is 0.99
  (determined by using the RM with the maximum $P$(RM)). The
  $\mathcal{R}$ spectra are sampled from -100 rad/m$^2$ to +100
  rad/m$^2$ in steps of 1 rad/m$^2$. In the bottom 2 panels, solid
  lines indicate the calculated $P(\mathcal{R})$\ and $\Phi(\mathcal{R})$\ spectra, and the dashed
  lines indicate the fitted RMSF.
}
\label{simple_spectrum}
\end{figure} 

\begin{figure}[t]
\resizebox{\hsize}{!}{\includegraphics[width=8.5cm]{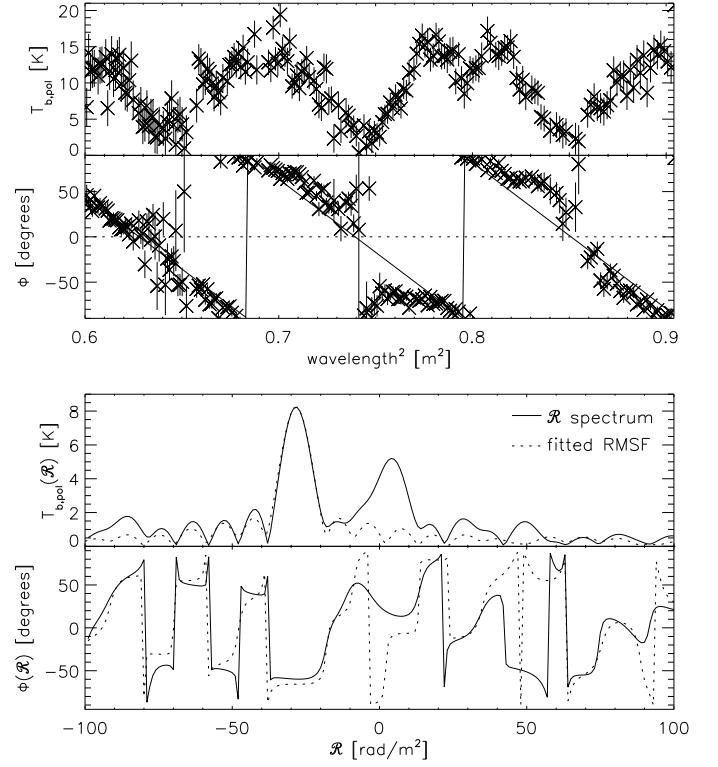}}
\caption{Identical to Fig. \ref{simple_spectrum}, but for a
  line of sight towards a polarized extragalactic source that also
  contains diffuse Galactic emission. The $\chi^2_{\mathrm{red}}$ of
  the RM fit is 8.72.  }
\label{l106m274}
\end{figure} 

In general the observations will not cover all wavelengths down to 0 m$^2$, which means that $\mathcal{R}$-extended structures will be missing from the $P(\mathcal{R})$\ spectra. This is similar to the missing large-scale structure problem interferometers suffer from when observing extended emission. 
If the source has a Gaussian $\vec{P}(\mathcal{R})$\ distribution along the line of sight, Brentjens \& De Bruyn calculate that in the reconstructed $\vec{P}(\mathcal{R})$\ spectrum $\mathcal{R}$ scales larger than

\begin{eqnarray}
\mathrm{maximum\ scale\ [radians/m^2]}\ \approx\ \frac{\pi}{\lambda^2_{\mathrm{min}}}
\label{ch3_maxscale.eqn}
\end{eqnarray}

\noindent
are suppressed by more than 50\% by the missing short wavelengths.

Analogous to how in an interferometer the size of the individual dishes sets the field of view, the channelwidth expressed in $\lambda^2$, $\delta\lambda^2$, determines the $\mathcal{R}$ for which the sensitivity has dropped to 50\% due to smearing of the polarization angles over individual channels, $\mathcal{R}_{\mathrm{max}}$: 

\begin{eqnarray}
|\mathcal{R}_{\mathrm{max}}|\ \mathrm{[radians/m^2]}\ =\ \frac{1.9}{\delta\lambda^2}
\label{ch3_rmax.eqn}
\end{eqnarray}

\noindent
. Also in this equation we replaced the $\sqrt{3}$ from the paper by Brentjens \& De Bruyn by 1.9 because this is closer to the correct value for the HWHM of the sinc RMSF response. Brentjens \& De Bruyn discuss the (minor) effect of using (possibly non-contiguous) channels with constant width in frequency instead of $\lambda^2$.

The addition of polarization vectors originating at different
distances along the line of sight in general leads to depolarization,
either because the vectors are emitted with different orientations
and/or because different amounts of Faraday rotation induce
misalignment of the polarization vectors. Faraday tomography can
separate the emission coming from subregions with different
$\mathcal{R}$ along the line of sight, thereby reducing the influence
of these sources of depolarization.

Brentjens \& De Bruyn (\cite{brentjensbruyn05}) note that one can also derotate the polarization angles of $\vec{P}({\mathcal{R}})$ to $\lambda^2 \ne 0$. In that case the $\lambda^2$ in the complex exponentials of Eqns. \ref{ch3_p_r.eqn} and \ref{ch3_rmsf.eqn} are replaced by ($\lambda^2-\lambda_0^2$). Brentjens \& De Bruyn showed that derotating to the weighted average of the observed $\lambda^2$,

\begin{eqnarray}
\lambda_0^2\ =\ \frac{\int\limits_{-\infty}^{\infty}W(\lambda^2)\lambda^2\mathrm{d}\lambda^2}{\int\limits_{-\infty}^{\infty}W(\lambda^2)\mathrm{d}\lambda^2}
\end{eqnarray}

\noindent
minimises the polarization angle variation over the main peak of the RMSF. However, now the polarization angle at the peak of the RMSF at Faraday depth $\mathcal{R}_{\mathrm{peak}}$ is no longer perpendicular to the magnetic field component in the plane of the sky. To determine the orientation of this magnetic field component, one needs to correct the polarization angles $\Phi$ by

\begin{eqnarray}
\Phi_0(\mathcal{R})\ =\ \Phi(\mathcal{R}) - \mathcal{R}_{\mathrm{peak}}\lambda_0^2
\end{eqnarray}

\noindent
. We derotated the polarization angles in the bottom panels of Figs. \ref{simple_spectrum} and \ref{l106m274} in this way.

\subsection{Examples}
We illustrate for a number of geometries of synchrotron-emitting and
Faraday-rotating regions the resulting $P(\lambda)$ and
$\Phi(\lambda)$, as well as the $\vec{P}(\mathcal{R})$ that is
reconstructed from the observations.

First, consider the case where there is only one infinitely thin source of emission along the line of sight, at Faraday depth $\mathcal{R_{\mathrm{1}}}$: $\vec{P}(\mathcal{R})\ =\ P_{\mathrm{1}}\ \delta(\mathcal{R}-\mathcal{R_{\mathrm{1}}})$, where $\delta(x)$ is the Dirac delta function. Then $\vec{P}(\lambda)$ = $P_{\mathrm{1}}\ \mathrm{e}^{2\mathrm{i}\mathcal{R_{\mathrm{1}}}\lambda^2}$: the $P(\lambda)$ are identical for all observing wavelengths $\lambda$, where we assumed that the synchrotron emission has a spectral index of 0. As Brentjens \& De Bruyn (\cite{brentjensbruyn05}) have discussed, a non-zero spectral index introduces only a distortion in the RMSF away from the main peak. 
$\Phi$ depends linearly on $\lambda^2$, and since the rotation measure RM $\equiv\ \partial\ \Phi /\partial\ \lambda^2$, RM = $\mathcal{R}$ in this case.
The $P(\mathcal{R})$ spectrum  that is reconstructed from the observed $P(\lambda)$ and $\Phi(\lambda)$ is identical to the original $P(\mathcal{R})$ spectrum, convolved with the RMSF. The situation described in this paragraph occurs in our data: in Fig. \ref{simple_spectrum} we show the measured $\vec{P}(\lambda^2)$ and reconstructed $\vec{P}(\mathcal{R})$ spectra for a line of sight with an essentially unresolved $P(\mathcal{R})$ spectrum.

Next, consider the case where there are two peaks of height $P_{\mathrm{1}}$ and $P_{\mathrm{2}}$ at Faraday depths $\mathcal{R}_{\ \mathrm{1}}$ and $\mathcal{R}_{\ \mathrm{2}}$. 
In this case $\vec{P}(\lambda)$ = $P_{\mathrm{1}}\mathrm{e}^{2\mathrm{i}\mathcal{R}_{\ \mathrm{1}}\lambda^2} + P_{\mathrm{2}}\mathrm{e}^{2\mathrm{i}\mathcal{R}_{\ \mathrm{2}}\lambda^2}$. This configuration produces a beat in $P(\lambda)$, because the 2 polarization vectors of length $P_{\mathrm{1}}$ and $P_{\mathrm{2}}$ rotate at different speeds $\mathcal{R}_{\ \mathrm{1}}$ and $\mathcal{R}_{\ \mathrm{2}}$ in the Stokes ($Q$,$U$) plane. Note that $\Phi$ no longer depends linearly on $\lambda^2$, which means that RM will not be constant. The $P(\mathcal{R})$ spectrum that is reconstructed from the observed $\vec{P}(\lambda)$\ will show the sum of 2 $\delta$-functions convolved with the RMSF. The peaks in the reconstructed $P(\mathcal{R})$\ spectrum will not lie at exactly $\mathcal{R}_{\ \mathrm{1}}$ and $\mathcal{R}_{\ \mathrm{2}}$, because the presence of the other peak will influence the shape of both peaks. Therefore, since $\Phi(\mathcal{R})$ is derived for the Faraday depth of the peak in the reconstructed $P(\mathcal{R})$ spectrum, the derived $\Phi(\mathcal{R})$ will not be the intrinsic position angle of the electric field of the emitted synchrotron radiation. 
When a polarized extragalactic point source is producing one of the 2 peaks, and the diffuse Galactic emission produces the second peak, the $\chi^2_{\mathrm{red}}$ of the linear fit of $\Phi$ versus $\lambda^2$ will be very high. In Fig. \ref{l106m274} we illustrate this case. The peak at $\mathcal{R}$ = -28 rad/m$^2$ is produced by an extragalactic source, and the region around $\mathcal{R}$ = +5 rad/m$^2$ is produced by Galactic emission in the foreground, as we established by comparing this line of sight to adjacent lines of sight, where the extragalactic source is not present.

The previous two examples dealt with infinitely thin emission
regions. If the emission region has a finite depth, but is not mixed
with Faraday-rotating ISM, the results we discussed in the previous
sections still hold. However, if the emitting and Faraday-rotating
regions are mixed, the $P(\mathcal{R})$\ distribution along the line of sight will
no longer be a $\delta$-function, and the reconstructed $P(\mathcal{R})$\ spectrum
will be wider than the width of the RMSF. If the synchrotron-emitting
and Faraday-rotating regions are fully mixed, and if the emissivity
and the amount of Faraday rotation per parsec are independent of the
position along the line of sight, then

\begin{eqnarray}
\vec{P}(\lambda)\ =\ P_{\mathrm{i}}\ \frac{\sin(\mathcal{R}_{\mathrm{max}}\lambda^2)}{\mathcal{R}_{\mathrm{max}}\lambda^2}\ \mathrm{e}^{2\mathrm{i}(\frac{1}{2}\mathcal{R}_{\mathrm{max}})\lambda^2}
\label{ch3_burn.eqn}
\end{eqnarray}

\noindent
where $\mathcal{R}_{\mathrm{max}}$ is the Faraday depth of the far side of the slab, and assuming that all radiation along the line of sight is emitted with a position angle of 0\degr. This result was first obtained by Burn (\cite{burn66}), and the distribution of synchrotron-emitting and Faraday-rotating regions along the line of sight that produces it has
become known as a `Burn slab'. Note that $\Phi$ again depends
linearly on $\lambda^2$, but that RM = 0.5$\mathcal{R}_{\ \mathrm{max}}$. Sokoloff et al. (\cite{sok98}) have shown that a linear
dependence of $\Phi$ on $\lambda^2$ occurs not only for a Burn slab, but for every 
distribution of synchrotron-emitting and Faraday-rotating regions that is symmetric
along the line of sight between the observer and $\mathcal{R}_{\ \mathrm{max}}$. The
reconstructed $P(\mathcal{R})$\ spectrum of a Burn slab has a finite extent, and may even
be wider than the width of a single RMSF. 

It is important to realise that for a general distribution of emitting and Faraday-rotating regions, only those structures show up in the $P(\mathcal{R})$\ spectrum that illuminate a column of Faraday-rotating ISM from the back. If Faraday rotation occurs also at larger distances, but is not illuminated from the back by synchrotron emission, this will not appear in the $\vec{P}(\mathcal{R})$\ spectrum.


\section{The data}\label{ch3_Sect: data_characteristics}

\begin{table*}[t]
\centering
\begin{tabular}{lcccccc}\hline\hline
Central position      & \multicolumn{6}{l}{ ($\alpha,\delta$)$_{\mathrm{2000}}$ = (109\degr,36.5\degr); ($l,b$) = (181\degr,20\degr) } \\
Mosaic size           & \multicolumn{6}{l}{9\degr $\times$ 9\degr}  \\
Pointings             & \multicolumn{6}{l}{7 $\times$ 7}\\
Frequencies           & \multicolumn{6}{l}{324--387 MHz} \\
                      & \multicolumn{6}{l}{202 independent frequency channels (Hamming taper)} \\
Resolution   & \multicolumn{6}{l}{2.76\arcmin $\times$ 4.70\arcmin} \\
Stokes $V$ noise level  & \multicolumn{6}{l}{6.2 mJy (2.0 K)} \\ 
$\mathcal{R}$\ noise level$^{\dagger}$ & \multicolumn{6}{l}{0.5 mJy (0.14 K)} \\ 
Conversion Jy--K$^{\ddagger}$ & \multicolumn{6}{l}{1 mJy/beam = 0.32 K (at 345 MHz)} \\ \hline
Shortest baseline (m)    & 36 & 48 & 60 & 72 & 84 & 96 \\
Observing date & \hspace{-1mm}03/01/17\hspace{-1mm} & \hspace{-1mm}03/01/21\hspace{-1mm} & \hspace{-1mm}02/12/18\hspace{-1mm} & \hspace{-1mm}02/12/05\hspace{-1mm} & \hspace{-1mm}03/01/22\hspace{-1mm} & \hspace{-1mm}03/01/13\hspace{-1mm} \\
\hspace{3mm} (yy/mm/dd) \\
Start time (UT) & 17:04 & 16:37 & 18:45 & 19:36 & 16:39 & 17:32 \\
End time (UT) & 05:04 & 04:36 & 06:43 & 07:34 & 04:37 & 05:31 \\ \hline
\end{tabular}
\caption[]{ Characteristics of the GEMINI data set.  Observing dates
  and times are given for each of the 12 hr observing runs, which have
  been indicated by their shortest baseline length. We calculated the
  conversion factor between mJy/beam and K at 345 MHz, the average of
  the $\lambda^2$ sampling of the (usable) frequency channels in our
  data set.\\ }
\label{ch3_mosaic_info.tab}
\end{table*}

The present data set was obtained with the WSRT, a 14-element E-W
interferometer of which 4 elements are moveable to improve (u,v)
coverage. Each of the telescope dishes is 25 m in diameter.  The
central coordinates of the region we study (in the constellation
Gemini) are $\alpha$ = $7^h 18^m$ and $\delta$ = 36\degr 24$\arcmin$
(J2000.0), which is $l\ \approx\ 181^{\circ}$ and $b\ \approx\ 20^{\circ}$
in Galactic coordinates.  The GEMINI region was observed in 6 12 hour
observing runs in December 2002 and January 2003 (see Table
\ref{ch3_mosaic_info.tab}). This yielded visibilities at baselines from 36 to
2760 meters, with an increment of 12 meters.  We tapered the
individual frequency channel maps in such a way that the synthesized
beamsize for all maps is that of the 385 MHz beam of 2.76\arcmin
$\times$ 4.70\arcmin\ (RA $\times$ DEC). Combining the 6 12 hr observing
runs puts the first grating ring at 4.1\degr\ (at 350 MHz) from the
pointing centre, outside the 3\degr$\times$3\degr\ area that we mapped
for each individual pointing.

\begin{figure*}[!t]
\centering
\resizebox{\hsize}{!}{\includegraphics[width=14cm]{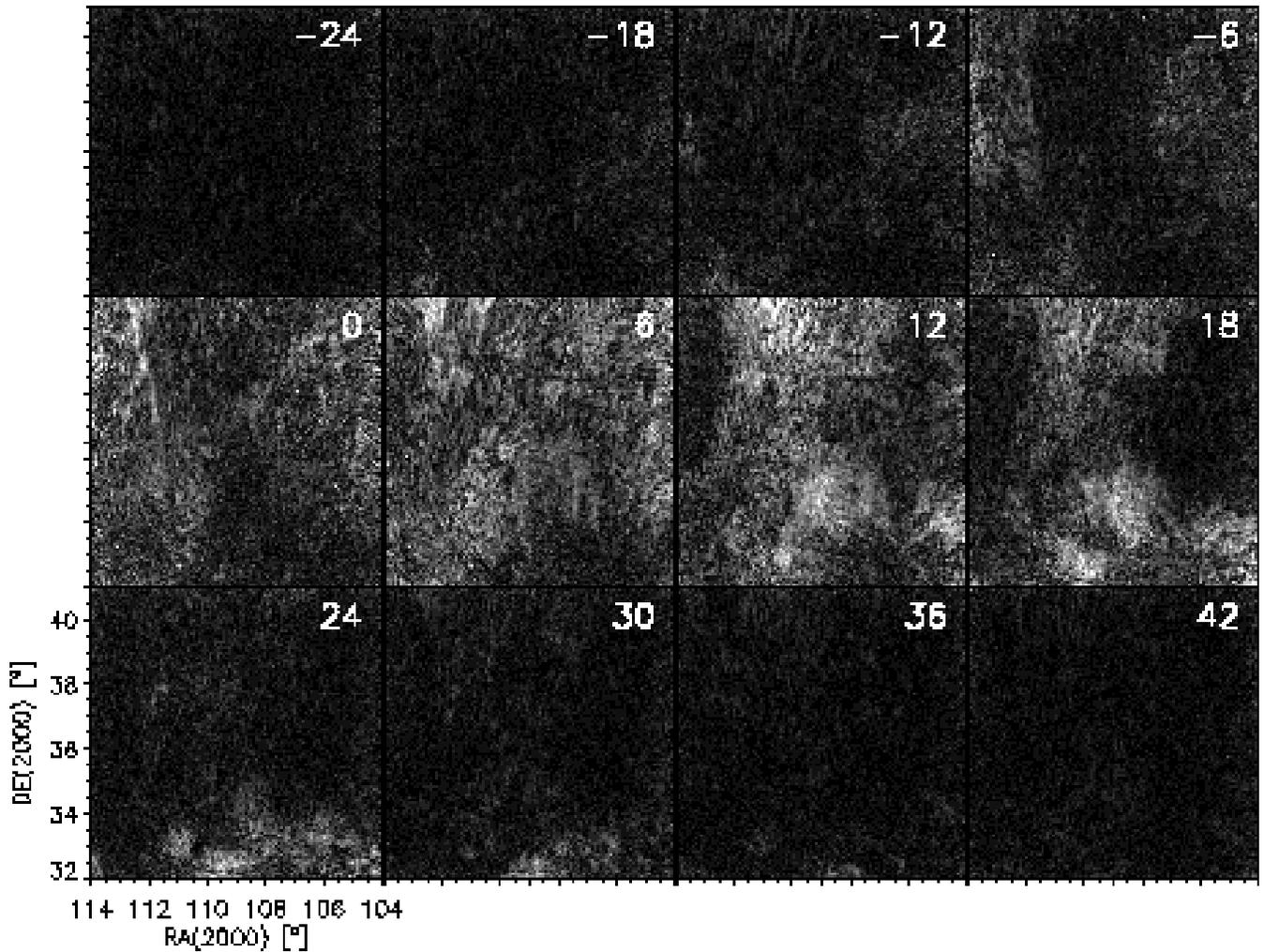}}
\caption{ Slices through the $P(\mathcal{R})$\ cube that show the strongest diffuse
  emission.  The $\mathcal{R}$\ of each slice is indicated in rad/m$^2$ in
  the upper right corner of each image. The RMSF has a FWHM of 12
  rad/m$^2$, and we sampled each RMSF with 2 images. The linear intensity
  scale is the same for all panels, and it saturates at 6.4 K.
}
\label{ch3_stamps.fig}
\end{figure*}

We mapped an area of about 9\degr$\times$9\degr\ with a 7$\times$7
pointing mosaic. In each night the same field was observed about 16
times, resulting in visibilities on 16 `spokes' in the (u,v) plane,
each time integrating for 40 seconds before moving to the next field.
The 1.2\degr\ distance between pointing centres suppresses off-axis
instrumental polarization to less than 1\% (Wieringa et
al. \cite{wieringa93}). In our analysis we leave out the edges of the
mosaic where instrumental polarization effects are not suppressed by
mosaicking.

The observations cover the frequency range between 324 and 387
MHz. 202 independent spectral channels are usable from a maximum of
224, each channel being 0.4 MHz wide (where we used a Hamming
taper). 10 channels were flagged because Stokes $V$ was contaminated by
radio frequency interference, and one channel was manually flagged in
all 6 12 hr runs.

The data were reduced using the $\mathtt{NEWSTAR}$ data reduction
package.  Dipole gains and phases and leakage corrections were
determined using the unpolarized calibrators 3C48, 3C147 and
3C295. The flux scales of both unpolarized and polarized calibrators
are set by the calibrated flux of 3C286 (26.93 Jy at 325 MHz - Baars
et al. \cite{baars77}). Due to an a-priori unknown phase offset
between the horizontal and vertical dipoles, signal can leak from
Stokes $U$ into Stokes $V$. We corrected for this by rotating the
polarization vector in the Stokes ($U$,$V$) plane back to the $U$
axis, assuming that there is no signal in $V$. The polarized
calibrator sources 3C345 and DA240 defined the sense of derotation
(i.e. to the positive or negative $U$-axis). Special care was taken to
avoid automatic flagging of real signal on the shortest baselines.
From Stokes $V$, which we assume to be empty, we estimate that the
average noise level in the mosaics of the individual channels is 6.2
mJy (2.0 K).  Instrumental polarization increases towards the edges of
the maps, therefore we excluded these regions in determining the noise
level.

These observations were carried out in the evening and at night to
limit solar interference and to reduce the importance of ionospheric
RM variations. From 2 lines of sight with a strong polarized signal
we estimate that the amounts of ionospheric Faraday rotation in the 6
nights are identical to within $\sim$ 10\degr, so we did not correct
for this. Ionosphere models indicate that the ionospheric contribution
to RM during the observing nights was only 0.6 rad/m$^2$
(Johnston-Hollitt, private communication).

An interferometer will not cover all baseline lengths down
to 0 m, which means that maps of the sky that were made using an
interferometer will miss structure on large angular scales. We
will return to this point in Sect. \ref{ch3_Sect: missing_lss}.

\section{Analysis}\label{ch3_Sect: analysis}
\subsection{The $\mathcal{R}$ datacube}\label{ch3_Sect: The curlyR datacube}
We calculated $\vec{P}(\mathcal{R})$ maps of the sky for Faraday depths from -1000 rad/m$^2$ to +998 rad/m$^2$ in steps of 6 rad/m$^2$. As the width of the RMSF along the $\mathcal{R}$ axis of the datacube is about 12 radians/m$^2$ for our data set, this gives Nyquist sampling in Faraday depth. Due to the finite channelwidth of the data, the sensitivity of the $P(\mathcal{R})$\ spectra will drop at large $\mathcal{R}$, as described by Eqn. \ref{ch3_rmax.eqn}. This becomes important for $|\mathcal{R}|\ \gtrsim\ 1.9/\mathrm{\delta\lambda^2}\ \approx\ 1250$ rad/m$^2$. 

\begin{figure}[!t]
\centering
\resizebox{\hsize}{!}{\includegraphics[width=12cm]{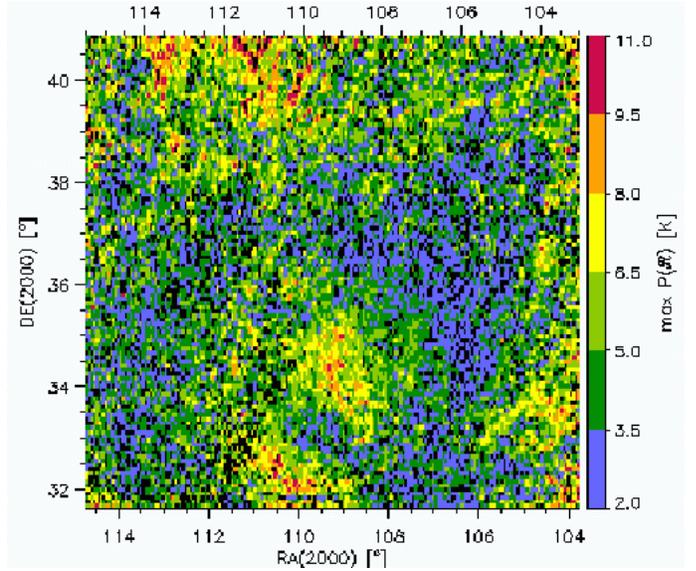}}
\caption{ $P(\mathcal{R})$ of the highest peak in the $\mathcal{R}$
  spectrum. Each pixel represents one telescope beam, and the pixel
  size is 2.8\arcmin$\times$4.9\arcmin. The lines of sight that are
  shown in this plot have a second peak in the $P(\mathcal{R})$\ spectrum that is at
  most half the strength of the main peak. Out of a total of 22.800
  lines of sight, 2720 do not obey this selection criterion, and they
  are shown in black, as are the 769 lines of sight that have
  $P(\mathcal{R}) < $ 2 K. The 1$\sigma$ noise level in the $P(\mathcal{R})$\ maps
  is 0.14 K. 
}
\label{ch3_p_cr_derot_lm.fig}
\end{figure}

\begin{figure}[!t]
\centering
\resizebox{\hsize}{!}{\includegraphics[width=12cm]{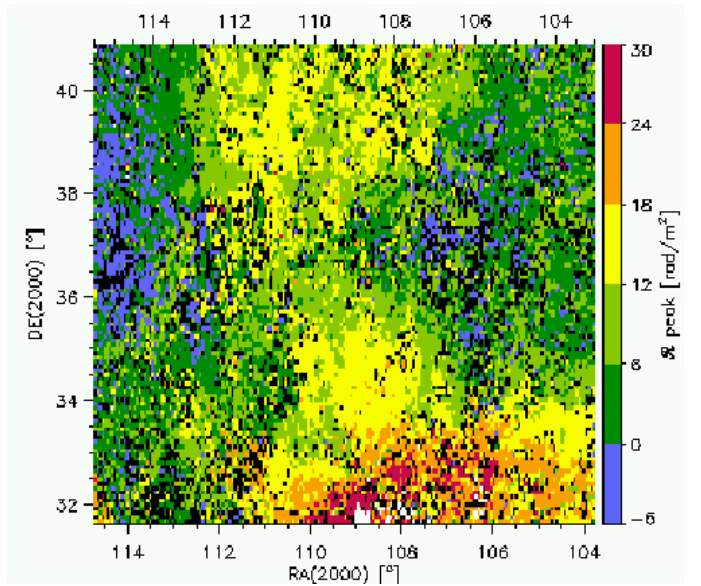}}
\caption{ $\mathcal{R}$ corresponding to the $P(\mathcal{R})$\ plotted in
  Fig. \ref{ch3_p_cr_derot_lm.fig}. The lines of sight plotted in these
  figures are selected by using the same criterion. Lines of sight
  indicated in white have $\mathcal{R} >$ 30 rad/m$^2$. Lines of sight
  indicated in black either did not pass the selection criterion or
  have $\mathcal{R} <$ -6 rad/m$^2$ (2720 and 727 lines of sight
  respectively). 
}
\label{ch3_cr_lm.fig}
\end{figure}

In Fig. \ref{ch3_stamps.fig} we show slices through our $\mathcal{R}$\ datacube that show strong Galactic emission. All of the images saturate at 6.4 K. 
In Figs. \ref{ch3_p_cr_derot_lm.fig} and \ref{ch3_cr_lm.fig} we summarise the information in the $\mathcal{R}$ datacube by plotting for each line of sight the maximum $P(\mathcal{R})$\ along that line of sight, and the $\mathcal{R}$\ at which this maximum occurs. We only plotted lines of sight where the main peak is more than twice as high as the second highest peak. From these figures it is clear that the majority of the lines of sight satisfies this condition. Because we treated the data in a slightly different way from what we described in Schnitzeler et al. (\cite{dominic07b}), there are some minor differences with the corresponding figures in that paper. From the $P(\mathcal{R})$\ maps at $|\mathcal{R}| > 200$ rad/m$^2$ we estimate that the noise level in the $P(\mathcal{R})$\ slices is 0.5 mJy (0.14 K). 

\begin{figure}[!t]
\centering
\resizebox{\hsize}{!}{\includegraphics[width=8.5cm]{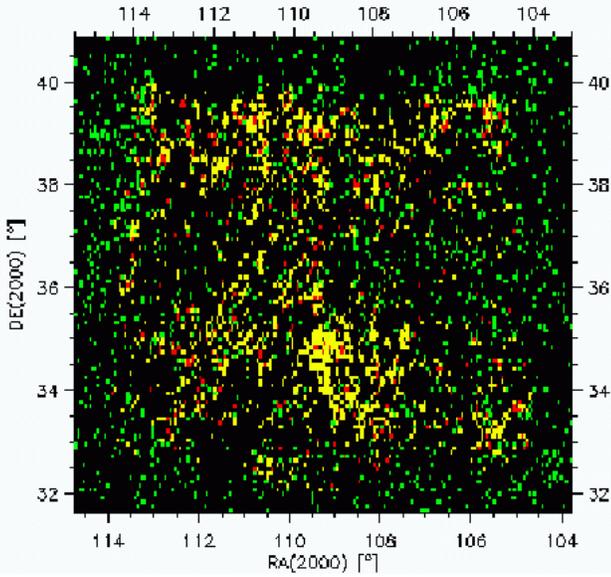}}
\caption{ Distribution of independent lines of sight with a main peak
  that satisfies the $\Delta$ criterion (green pixels), or with
  $\chi^2_{\mathrm{red}} < 2$ and max($P(\mathcal{R})$) $>$ 2 K
  (yellow pixels), or that satisfy all three criteria (red
  pixels). Lines of sight that did not pass these criteria are shown
  as black pixels. The outer edges of the mosaic show an increase in
  $\chi^2_{\mathrm{red}}$, which explains the absence of yellow
  pixels. 
}
\label{ch3_fscreens.fig}
\end{figure}

The complexity of a line of sight depends on whether the main peak is resolved, and on how strong second and higher-order peaks are. 
The $\Delta$ criterion that we introduced in Schnitzeler et al. (\cite{dominic07b}) can be used to address the issue of resolution. We defined $\Delta$ as the root-mean-square vertical separation in the $\mathcal{R}$ spectrum between the main peak in the $P(\mathcal{R})$\ spectrum and the best-fitting RMSF, and we compare the observed $\Delta$ to the distribution of $\Delta$ that we simulated for an input signal of the appropriate strength + noise. In Fig. \ref{ch3_fscreens.fig} we plot the lines of sight that have an unresolved main peak as green pixels. `Unresolved' means that the $\Delta$ value of the main peak is $\le$ the $\Delta$ that we found for 99\% of our simulations of signal + noise ($\approx$ the 3$\sigma$ level of the Rayleigh distribution). 

Lines of sight with linear $\Phi(\lambda^2)$\ relations have special geometries of Faraday-rotating and synchrotron-emitting regions, as discussed in Sect. \ref{ch3_Sect: Faraday tomography}.
We therefore fitted a straight line to our $\Phi(\lambda^2)$\ datapoints, where we included the periodicity of the data by using the procedure described in Schnitzeler et al. (\cite{dominic07b}). Lines of sight with $\chi^2_{\mathrm{red}}$ $<$ 2 are indicated in  Fig. \ref{ch3_fscreens.fig} as yellow pixels, and lines of sight that have both a not measurably resolved main peak and a low $\chi^2_{\mathrm{red}}$\ are indicated as red pixels. Since many lines of sight have a low $\chi^2_{\mathrm{red}}$, but a high $\Delta$, there are many yellow pixels, but relatively few red pixels.

We simulated the $P(\mathcal{R})$\ response from the main peak plus noise by calculating the Fourier transform from Eqn. \ref{ch3_p_r.eqn} for an input signal of a given strength + the noise levels that we derived from the Stokes $V$ maps of the individual frequency channels. We assume that the Stokes $V$ maps contain no signal. We ran our simulations for a grid of input signal strengths, that was matched to the strengths of the main peak in the $P(\mathcal{R})$\ spectra that we encountered in the data. In this way we calculated for each $\mathcal{R}$\ in the spectrum the level below which  99\% of the RMSF plus noise simulations lie. 

The main peak in the $P(\mathcal{R})$\ spectrum is not measurably resolved for about 8\% of the lines of sight in the current data set, in the sense that its value of $\Delta$ is lower than the $\Delta$ that we find for 99\% of our simulations of a RMSF + noise. This value is lower than the 14\% quoted in Schnitzeler et al. (\cite{dominic07b}). This difference might be produced because we now use frequency channel maps that all have been tapered to the same beamsize. A not measurably resolved peak means that a Faraday-rotating region is physically separated from a region with synchrotron emission; mixing of the 2 regions produces broad peaks in the $P(\mathcal{R})$\ spectrum. Since we only tested the main peak in our $P(\mathcal{R})$\ spectra, and no other features in the $P(\mathcal{R})$\ spectra, emission regions with a Faraday screen in front of them will be even more numerous than in 8\% of the lines of sight. RFI and calibration errors can only increase $\Delta$, which is another reason why there are probably more Faraday screens present in the ISM. 

Intuitively one would not expect such a separation between the emitting and Faraday-rotating regions along the line of sight: cosmic rays that produce synchrotron radiation are expected to be present wherever there is a magnetic field. Therefore one possibility could be that we simply do not see the synchrotron emission because the full magnetic field vector is pointing along the line of sight. The magnetic field component perpendicular to the line of sight, which determines the synchrotron emissivity, is in that case 0. Such a region would from our point of view then only produce Faraday rotation. Even though the strengths of the magnetic field component along the line of sight that we determine in Sect. \ref{ch3_Sect: magnetic field} are low, these are averages over the line of sight, and they are biased towards regions where the thermal (i.e. 10$^4$ K) electron density is higher. 

The majority of lines of sight in our data have a secondary peak that is higher than what would be expected for one in 10$^4$ noise realisations. However, the main peak is typically more than twice as high as the secondary peak, and thus dominates most of the lines of sight in the GEMINI region. This makes these lines of sight relatively simple, and we did not deconvolve the $P(\mathcal{R})$\ spectra. The reason why these lines of sight are not very complex could be because we are looking in the direction of the Galactic anti-centre, at an intermediate Galactic latitude. 


\subsection{Missing large-scale structure}\label{ch3_Sect: missing_lss}

Structure on large angular scales is missing from interferometric observations if information on short baselines is not available. Differences in Faraday rotation in between the source of the polarized emission and the observer can for neighbouring lines of sight `transfer' polarized signal from large angular scales to small enough angular scales so that it can be picked up by the interferometer. Since this effect does not operate on Stokes $I$, this is the canonical explanation why structure in polarized intensity does not appear to have a counterpart in total intensity. In this section we expand on this idea, and we show that modulation by a linear gradient in Faraday depth $\mathcal{R}$\ shifts the angular frequency spectrum as a whole.

For a linear variation of $\mathcal{R}$ with position $l$ (i.e. a fixed gradient $\nabla{\mathcal{R}}$), and wavelength $\lambda$, this modulation of the polarization vector $\vec{P}$ can be expressed as

\begin{eqnarray}
\vec{P}_{\mathrm{obs}}(l)\ =\ \vec{P}_{\mathrm{em}}(l)\ \mathrm{e}^{2\mathrm{i}\nabla{\mathcal{R}}\lambda^2 l}
\end{eqnarray}

\noindent
where subscripts `em' and `obs' refer to the emission before and after it gets modulated by the foreground gradient in $\mathcal{R}$. This linear modulation with position shifts the angular frequency spectrum as a whole:

\begin{eqnarray}
\mathrm{FT}(\vec{P}_{\mathrm{obs}})(u) & = &
  \int_{-\infty}^{\infty} \mathrm{d}l\ \vec {P}_{\mathrm{obs}}(l)\ \mathrm{e}^{-2\pi \mathrm{i}ul} \\
  & = & 
  \int_{-\infty}^{\infty} \mathrm{d}l\ \vec {P}_{\mathrm{em}}(l)\ \mathrm{e}^{2\mathrm{i}\nabla{\mathcal{R}}\lambda^2l} \mathrm{e}^{-2\pi \mathrm{i}ul} \nonumber \\
  & = &\int_{-\infty}^{\infty} \mathrm{d}l\ \vec {P}_{\mathrm{em}}(l)\ \mathrm{e}^{-2\pi \mathrm{i}(u-\frac{1}{\pi}\nabla{\mathcal{R}}\lambda^2)\ l} \nonumber \\
  & = & \mathrm{FT}(\vec{P}_{\mathrm{em}})(u-\frac{1}{\pi}\nabla{\mathcal{R}}\lambda^2) 
\label{ch3_freq_shift.eqn}
\end{eqnarray}


\noindent
. One consequence of this is that foreground Faraday modulation makes the 0-angular frequency from $\vec{P}_{\mathrm{em}}(l)$ visible if $|\nabla\mathcal{R}|$ is large enough. To make it detectable at the 25 m (projected) baseline length, which can be obtained with the WSRT, requires a $|\nabla{\mathcal{R}}|\ \approx\ 6$ rad/m$^2$/field of view at 350 MHz (where 1 field of view is 3\degr\ wide for our observations). Since we observe gradients of this magnitude in the individual pointings of the GEMINI data set (Fig. \ref{ch3_cr_lm.fig}), Fig. \ref{ch3_p_cr_derot_lm.fig} and Fig. \ref{ch3_cr_lm.fig} contain at least the 0-angular frequency = total polarized intensity component of $\vec{P}_{\mathrm{em}}$.

The direction of the shift in Eqn. \ref{ch3_freq_shift.eqn} is set by the sign of $\nabla\mathcal{R}$. Since the WSRT only measures one half of the (u,v)-plane, this means that the angular frequency spectrum could move away from the measurement points in the (u,v)-plane. However, $\vec{P}_{\mathrm{obs}}(l)=Q_{\mathrm{obs}}(l)+\mathrm{i}U_{\mathrm{obs}}(l)$, and since $Q_{\mathrm{obs}}(l)$ and $U_{\mathrm{obs}}(l)$ are real quantities, their Fourier transforms are hermitian. The gradient from Eqn. \ref{ch3_freq_shift.eqn} always moves both the positive and the negative angular frequencies of $Q_{\mathrm{em}}(l)$ and $U_{\mathrm{em}}(l)$ towards or away from the origin in the (u,v)-plane. Therefore, $\vec{P}_{\mathrm{obs}}(l)$ can be reconstructed even if only one half of the (u,v)-plane is observed. 

Note that foreground Faraday modulation shifts some angular frequencies from $\vec{P}_{\mathrm{em}}(l)$ towards smaller (in an absolute sense) angular frequencies. At some point these angular frequencies from $\vec{P}_{\mathrm{em}}(l)$ will no longer be visible for an interferometer, because it can only measure visibilities from a certain shortest baseline. For the analysis that we present in the remainder of this article it is sufficient that our data contain the total polarized intensity component of $\vec{P}_{\mathrm{em}}$. Single-dish observations are therefore still required to produce maps that contain all angular frequencies.

If Faraday rotation in the foreground produces strong variations in $\mathcal{R}$ over the telescope beam, then this will also lead to beam depolarization. For a Gaussian beam, a gradient of more than 12 radians/m$^2$/field of view is required to give more than 10\% beam depolarization (Sokoloff et al. \cite{sok98}), and such steep gradients are not observed in the GEMINI data.

As a final note in this section, we showed in Schnitzeler et al. (\cite{dominic07a}) that the polarization angle gradients in the WENSS data in the region 130\degr\ $< l <$ 170\degr, -5\degr\ $< b <$ 30\degr\ are on average about 2 radians/m$^2$/\degr\ along Galactic latitude, which translates into the 6 radians/m$^2$/field of view required to shift the angular frequency spectrum towards observable frequencies. These WENSS gradients therefore also show at least the total polarized intensity component of the angular frequency spectrum. 

\section{The line-of-sight component of the large-scale magnetic field}\label{ch3_Sect: magnetic field}
In this section we derive the strength of the magnetic field component along the line of sight, $B_{\|}$. We use the Faraday depths from Fig. \ref{ch3_cr_lm.fig}, and we determine the contribution from thermal electrons to $\mathcal{R}$ (Eqn. \ref{ch3_curlyR.eqn}) by converting H$\alpha$ intensities from the WHAM survey (Haffner et al. \cite{haffner03}) first to emission measures EM = $\int_{0}^{\infty} n^2_{\mathrm{e}}\mathrm{d}l$, and then to dispersion measures DM = $\int_{0}^{\infty} n_{\mathrm{e}}\mathrm{d}l$, using an empirical relation between EM and DM that Berkhuijsen et al. (\cite{berkhuijsen06}) derived for Galactic pulsars. The integrals for EM and DM are over the entire line of sight, $[$EM$]$ = cm$^{-6}$pc, $[$DM$]$ = cm$^{-3}$pc, $[n_{\mathrm{e}}]$ = cm$^{-3}$ and [d$l$] = pc. 

To estimate DM, we could also have used the electron density models by Reynolds (\cite{reynolds91}) or by Cordes \& Lazio (\cite{cordeslazio03}) as alternatives to the relation between EM and DM that Berkhuijsen et al. derived. In the Reynolds model, 40\% of the line of sight contains free electrons, at a density of 0.08 cm$^{-3}$, and there are no free electrons in the remainder of the line of sight. Cordes \& Lazio fitted a smooth electron density model consisting of a thin and thick disc + spiral arms to the DM measured for about 1200 Galactic pulsars and 100 extragalactic sources. Small-scale structure in the Galactic ISM is included in their model as a list of local clumps and voids. Berkhuijsen et al. used DM measured for a sample of 157 pulsars, and accurately determined the EM between us and the pulsar by correcting for the EM that is produced beyond the pulsar, and for reddening occurring in front of the pulsar. By combining Berkhuijsen et al.'s statistical description of the ISM with the WHAM H$\alpha$ intensities that are measured on a $\approx$ $1^{\circ}\times 1^{\circ}$ grid, we can more accurately determine DM in specific directions than is possible with the models by Reynolds or by Cordes \& Lazio.

Significant numbers of free electrons are present in both the warm ionized medium (WIM) and hot intercloud medium (HIM) phases of the ISM. 
Snowden et al. (\cite{snowden97}) modelled the Galactic X-ray halo from ROSAT diffuse X-ray background maps. Their model consists of a cylindrical bulge of radius 5.6 kpc, with a HIM electron density that decays exponentially away from the Galactic midplane. The HIM electron density in their model is in the Galactic midplane a factor of 10 smaller than the average WIM electron density (3.5 $\times$ 10$^{-4}\ \mathrm{cm}^{-3}$ compared to $\sim$ 3 $\times 10^{-3}\ \mathrm{cm}^{-3}$). If we assume that the HIM electron densities in the outer Galaxy (in the direction of GEMINI) are comparable to the densities in the inner Galaxy, then the DM produced in the HIM is small compared to the DM of the WIM, and we therefore neglected the former.

\begin{figure}[!t]
\centering
\resizebox{\hsize}{!}{\includegraphics[width=12cm]{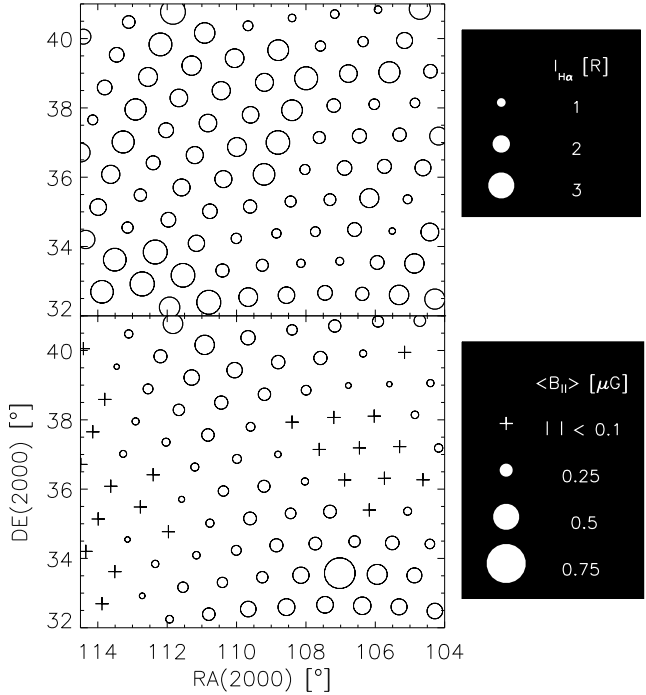}}
\caption{WHAM H$\alpha$ intensities in Rayleigh (top panel) and the
  electron-density weighted strengths of the magnetic field component
  parallel to the line of sight $\langle B_{\|}\rangle$ [$\mu$G] that
  we derived (bottom panel). Open circles indicate positive $\langle
  B_{\|}\rangle$; there are no negative $\langle B_{\|}\rangle < $
  -0.1 $\mu$G in this figure. $|\langle B_{\|}\rangle| < $ 0.1 $\mu$G
  are shown as `+'.  }
\label{ch3_wham.fig}
\end{figure}

In the top panel of Fig. \ref{ch3_wham.fig} we plot the H$\alpha$ intensities (in Rayleigh) for WHAM lines of sight in the GEMINI region. To convert the H$\alpha$ intensities to EM, we use Eqn. 1 from Haffner et al. (\cite{haffner98}):

\begin{equation}
\mathrm{EM}=2.75 T_4^{0.9}\ I_{\mathrm{H}\alpha}\ \mathrm{e}^{\mathrm{2.2\ E(B-V)}}
\label{wham2em}
\end{equation}

\noindent
. $T_4$ is the temperature of the WIM gas in units of 10$^4$ K, which is
typically 0.8 (Reynolds \cite{reynolds85}). $I_{\mathrm{H}\alpha}$ is
the H$\alpha$ intensity in Rayleigh, and E(B-V) is the interstellar
B-V reddening. We determined EM for each of the WHAM lines of sight by using reddening values from Schlegel et al. (\cite{schlegel98}) and the programs available on Marc Davis' website\footnote{\tt{http://astro.berkeley.edu/$\sim$marc/dust/data/data.html}}. 

The 157 pulsars with $|b|> 5^{\circ}$ in the sample by Berkhuijsen et al. follow the relation

\begin{equation}
\mathrm{EM}= (0.042 \pm 0.014)\ \mathrm{DM}^{1.47 \pm 0.09}
\label{em2dm}
\end{equation}

\noindent
. With the $\mathcal{R}$ from Fig. \ref{ch3_cr_lm.fig} and the DM that we calculated from the WHAM H$\alpha$ intensities, we can then calculate the electron-density weighted average $B_{\|}$ along the line of sight, $\langle B_{\|}\rangle$:

\begin{equation}
\langle B_{\|}\rangle \equiv \frac{\int\limits_{\mathrm{line\ of\ sight}}^{\mathrm{observer}} n_e\vec{B}\cdot\mbox{d}\vec{l}}{\int\limits_{\mathrm{line\ of\ sight}} n_e \mbox{d}l} = \frac{\mathcal{R}}{0.81\ \mathrm{DM}}
\label{avbparr}
\end{equation}

\noindent
where $\langle B_{\|}\rangle$ is in $\mu$G. In the bottom panel of Fig. \ref{ch3_wham.fig} we show the $\langle B_{\|}\rangle$ that we calculated in this way. Since the WHAM beam (FWHM $\approx$ 1\degr) is much larger than the synthesized WSRT beam, we used the $P(\mathcal{R})$-weighted $\mathcal{R}$ average over the WHAM beam to calculate $\langle B_{\|}\rangle$. Also note that since Schlegel et al. give reddening values for the entire line of sight through the Galaxy, the EM and DM that we calculate are also values for the entire line of sight. Since we do not know where the polarized emission that we observe originates, the $|\langle B_{\|}\rangle|$ that we derive from Eqn. \ref{avbparr} are lower limits to the actual $|\langle B_{\|}\rangle|$. 

An upper limit for $B_{\|}$ can be estimated by using a pitch angle of 8\degr\, and a strength of 4 $\mu$G for the large-scale magnetic field (both values from Beck \cite{beck07a}), which gives $B_{\mathrm{reg},\|,\mathrm{max}} \approx 0.6$ $\mu$G. In the presence of a random magnetic field,  $B_{\|}$ approaches the $B_{\mathrm{reg},\|}$ component of the regular field if either the reversals in the direction of the random field are unimportant, or if the line of sight is long enough so that the contributions to $\mathcal{R}$ from such small-scale reversals cancel each other along the line of sight. As discussed in Haverkorn et al. (\cite{haverkorn04}), the random magnetic field has a correlation length of order 10 parsec. Shortly we will estimate that the length of the line of sight through ionized regions in the direction of GEMINI is of the order of 130 parsec. The number of `draws' of the random field is therefore small, which means that the random magnetic field will have a non-negligible contribution (compared to the large-scale field) to the $\langle B_{\|}\rangle$ that we derive. The $\langle B_{\|}\rangle$ in the bottom panel of Fig. \ref{ch3_wham.fig} already show that the $\langle B_{\|}\rangle$ in GEMINI vary much more than what we would expect on the basis of the large-scale magnetic field alone.

The $\langle B_{\|}\rangle$ in the bottom panel of Fig. \ref{ch3_wham.fig} are smaller than or comparable to the $B_{\mathrm{reg},\|,\mathrm{max}}$ that we estimated. The $\langle B_{\|}\rangle$ we derive are small, which can be expected since we are looking in the direction of the Galactic anti-centre, where the regular magnetic field is mostly perpendicular to the line of sight. It is clear from the bottom panel in Fig. \ref{ch3_wham.fig} that variations in EM alone are not enough to explain the measured variation in $\mathcal{R}$, but that variations in $B_{\|}$ are required as well. Further support for this is given by the fact that $\mathcal{R}$ in Fig. \ref{ch3_cr_lm.fig} changes sign in the area covered by the GEMINI data, which can only be explained by a change in the direction of $B_{\|}$. The $B_{\|}$ structure  in Fig. \ref{ch3_cr_lm.fig} is not associated with any of the 3 largest non-thermal loops in the Galaxy, the North Polar Spur, Cetus Arc and Loop III. The locations of these structures are described in Elliott (\cite{elliott70}).

Berkhuijsen et al. also constructed a model of the electron density distribution along the line of sight, that we can use to estimate the length of the line of sight that produces the observed EM. Similar to the model by Reynolds (\cite{reynolds91}), their model consists of cells that are empty or filled with a constant electron density. Contrary to the Reynolds model, the electron density of the filled cells (`clouds'), $n_{\mathrm{e,cloud}}$, is allowed to vary between lines of sight:

\begin{equation}
n_{\mathrm{e,cloud}} = \frac{\mathrm{EM}}{\mathrm{DM}} = 0.12\ \mathrm{EM}^{\ 0.32}
\end{equation}

\noindent
where we used Eqn. \ref{em2dm}. Berkhuijsen et al. found that the filling factor $f$ of electron clouds depends inversely on the electron density in the clouds: $f= (0.0184\pm 0.0011)\ n_{\mathrm{e,cloud}}^{-1.07\pm0.03}$. We can use this relation to determine the length of the line of sight $D$ (in parsec) that produces the observed EM; note that $D$ includes both the fraction of the line of sight that passes through the electron clouds and also the electron-free regions between the clouds:

\begin{equation}
D = \frac{1}{f}\left(\frac{\mathrm{DM}^2}{\mathrm{EM}}\right) =  404\ \mathrm{EM}^{\ 0.70}
\end{equation}


\noindent
. E(B-V) = 0.071 magnitudes for ($\alpha,\delta$) = (110$^{\circ}$,37$^{\circ}$), which means that the average H$\alpha$ intensity in GEMINI of 1.78 R translates into an EM = 4.68 cm$^{-6}$pc. This gives $n_{\mathrm{e,cloud}}$ = 0.19 cm$^{-3}$, $f$ = 0.11, and $D$ = 1.2 kpc. This filling factor and $D$ imply that the combined pathlength through the electron clouds is $f\times D \approx$ 130 parsec. We calculated EM and DM for the entire line of sight, but it is unclear whether we can observe the polarized emission coming from the furthest parts along the line of sight. $D$ is therefore an upper limit to the length of the line of sight that produces the $\mathcal{R}$ in Fig. \ref{ch3_cr_lm.fig}.

\section{Polarized point sources}\label{ch3_Sect: point sources}
%
%

\begin{figure}[!t]
\centering
\resizebox{\hsize}{!}{\includegraphics[width=12cm]{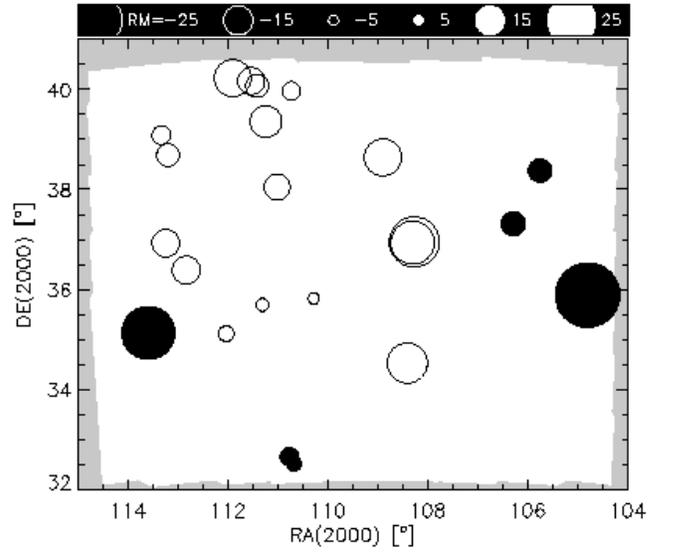}}
\caption{ RM distribution for the polarized point sources from table
  \ref{ch3_egs.tab}. The size of the circles is proportional to RM, shown in
  the scale on top of the figure in units of radians/m$^2$. Black
  circles indicate negative RM, and white circles indicate positive
  RM. $|$RM$|$ $\lesssim\ 4$ rad/m$^2$\ are missing from this figure
  because of our selection criteria. We excluded the grey region when
  looking for polarized point sources, because instrumental
  polarization levels are higher there.
}
\label{ch3_cr_egs.fig}
\end{figure}

\begin{table}[t]
\centering
\begin{tabular}{crcccc}\hline\hline
\hspace{-1mm}$($RA, DEC$)$ & \multicolumn{1}{c}{\hspace{3mm}RM} & $\chi^2_{\mathrm{red}}$ & $P({\mathcal{R}})_{\mathrm{max}}$ & \multicolumn{1}{c}{$I$} & $P({\mathcal{R}})_{\mathrm{max}}$/$I$\\ 
 \hspace{-1mm}$($\degr,\degr$)$ & $[$rad/m$^2$$]$ & & $[$mJy$]$ & $[$mJy$]$ &  \% \\ \hline
$($113.60, 35.12$)$ & -27.6 $\pm$  0.1 &   0.5 &  23.4 &  329 &  7.1\\
$($113.25, 36.92$)$ &  14.2 $\pm$  0.3 &   0.6 &   6.1 &  177 &  3.4\\
$($113.34, 39.09$)$ &   9.6 $\pm$  0.4 &   0.7 &   7.9 &  236 &  3.3\\
$($113.20, 38.69$)$ &  11.7 $\pm$  0.4 &   0.8 &   5.8 &  184 &  3.2\\
$($112.84, 36.38$)$ &  14.3 $\pm$  0.5 &   1.1 &   6.9 &  337 &  2.0\\
$($112.04, 35.11$)$ &   8.7 $\pm$  0.4 &   0.8 &   4.6 &  214 &  2.2\\
$($111.91, 40.23$)$ &  18.9 $\pm$  0.4 &   1.1 &   6.2 &  151 &  4.1\\
$($111.54, 40.17$)$ &  13.8 $\pm$  0.5 &   1.4 &   5.7 &  272 &  2.1\\
$($111.32, 35.69$)$ &   6.5 $\pm$  0.3 &   0.5 &  10.9 &  277 &  3.9\\
$($111.42, 40.07$)$ &  11.8 $\pm$  0.3 &   0.9 &   5.3 &  152 &  3.5\\
$($111.24, 39.36$)$ &  16.3 $\pm$  0.6 &   1.4 &   7.0 &  333 &  2.1\\
$($111.02, 38.05$)$ &  13.3 $\pm$  0.2 &   0.5 &  10.0 &  375 &  2.7\\
$($110.77, 32.65$)$ &  -9.8 $\pm$  0.7 &   2.8 &   5.4 &  232 &  2.3\\
$($110.68, 32.51$)$ &  -7.8 $\pm$  0.6 &   2.1 &   4.8 &  242 &  2.0\\
$($110.74, 39.97$)$ &   9.4 $\pm$  0.3 &   0.6 &  11.8 &  239 &  4.9\\
$($110.29, 35.82$)$ &   5.7 $\pm$  0.5 &   0.9 &   6.3 &  258 &  2.4\\
$($108.90, 38.64$)$ &  19.0 $\pm$  0.4 &   1.9 &   4.5 &  197 &  2.3\\
$($108.42, 34.52$)$ &  20.5 $\pm$  0.2 &   1.0 &  17.4 &  828 &  2.1\\
$($108.31, 36.93$)$ &  22.2 $\pm$  0.2 &   1.0 &   5.3 &  214 &  2.5\\
$($108.27, 36.95$)$ &  25.7 $\pm$  0.6 &   2.7 &   4.4 &  210 &  2.1\\
$($106.30, 37.31$)$ & -12.7 $\pm$  0.4 &   1.2 &   4.4 &  195 &  2.3\\
$($105.76, 38.38$)$ & -12.3 $\pm$  0.6 &   1.4 &   3.1 &  150 &  2.0\\
$($104.81, 35.88$)$ & -33.4 $\pm$  0.4 &   1.6 &  15.1 &  625 &  2.4\\ \hline
\end{tabular}
\caption[]{ Properties of the polarized point sources we found in
  our data. Shown are the equatorial coordinates (J2000.0) of the
  source in decimal degrees, the RM that we fitted and its error
  (both in rad/m$^2$), the reduced $\chi^2$ ($\chi^2_{\mathrm{red}}$)
  of the fit, the maximum $P(\mathcal{R})$\ along the line of sight and the total
  intensity $I$ (both in mJy), and the polarization fraction (in \%).
} 
\label{ch3_egs.tab}
\end{table}

Since the line of sight towards extragalactic sources passes through the entire  Milky Way, whereas different depolarization effects can limit the line of sight of the diffuse emission, the information we obtain from the $\mathcal{R}$ of extragalactic sources can complement what we learn from the diffuse emission. However, extragalactic sources will also have an intrinsic $\mathcal{R}$. 
%
%
%

To find polarized point sources, we first made maps that only contain baselines $>$ 250 m, to filter out the strong diffuse emission from our $P(\mathcal{R})$\ datacube. These maps use a Gaussian taper that reaches a value of 0.25 at a baseline length of 2500 m. We then Fourier transformed these channel maps from $\lambda^2$ space to $\mathcal{R}$ space, and constructed $P(\mathcal{R})$\ maps from $\mathcal{R}$\ = -1000 rad/m$^2$\ to $\mathcal{R}$\ = +998 rad/m$^2$, in a way identical to what we described in Sect. \ref{ch3_Sect: The curlyR datacube}.

The polarized intensity of an extragalactic source with rotation measure RM will be attenuated by bandwidth depolarization according to sinc($|$RM$|\delta\lambda^2$), if the polarization angle varies linearly with $\lambda^2$. Here $\delta\lambda^2$ is the $\lambda^2$ width of the frequency channels. The channels that we use have on average a $\delta\lambda^2\ \approx\ 1.5\times10^{-3}$ m$^2$, therefore bandwidth depolarization of a point source with $|\mathcal{R}|$ = 1000 rad/m$^2$\ will reduce the signal to 66\%, which would still be detectable.

We looked for each line of sight in the $P(\mathcal{R})$\ datacube if 1) there was a Stokes $I$ counterpart brighter than 150 mJy,\ 2) if the maximum $P(\mathcal{R})$\ along the line of sight was more than a 5$\sigma$ detection and 3) if the ratio of the maximum $P(\mathcal{R})$\ along the line of sight/ total intensity $I$ was larger than 0.02, meaning that this source is more than 2\% polarized (the instrumental polarization level is about 1\%). We also exclude sources with $|\mathcal{R}| \le 4$ rad/m$^2$\ because instrumentally polarized point sources will show up in the $P(\mathcal{R})$\ datacube at 0 rad/m$^2$, and the finite width of the RMSF along the $P(\mathcal{R})$\ axis will produce a strong signal also in the vicinity of 0 rad/m$^2$. Furthermore we excluded a region of about 0.9\degr\ from the edge of the mosaic, where there is no overlap between pointings, and instrumental polarization levels are therefore much higher there. We then fitted a RM to the $\Phi(\lambda^2)$ distribution of the sources that satisfy these criteria in the way described in Schnitzeler et al. (\cite{dominic07b}).


In Table \ref{ch3_egs.tab} we list the properties of the sources that we found, and we plot the RM of these sources in Fig. \ref{ch3_cr_egs.fig}. The low $\chi^2_{\mathrm{red}}$ of the RM fits indicate that there is only one $P(\mathcal{R})$\ component along the $\mathcal{R}$ axis, and that we have effectively filtered out the diffuse Galactic contribution to the $P(\mathcal{R})$\ spectrum. This, in combination with the fact that these sources are detected in maps where we left out low angular frequencies on the sky, means that the sources that we detected are compact in all 3 dimensions of the $P(\mathcal{R})$\ cube, and are therefore likely to be either polarized pulsars or polarized extragalactic sources. Since the surface density of the latter is so much higher than that of the former, the sources that we detected are probably extragalactic in origin.

Most of the sources in the upper left quadrant of Fig. \ref{ch3_cr_egs.fig} show positive RMs of $\approx$ 5--15 rad/m$^2$, which is similar to the Faraday depths of the diffuse emission from Fig. \ref{ch3_cr_lm.fig}. In other regions the source density is insufficient to reach definite conclusions. The RM that we observe for an extragalactic source is the combination of a Galactic RM and a RM that is intrinsic to the source. However, since a number of sources show the same RM, and since the intrinsic RM of these sources are uncorrelated, what we see is probably mostly the Galactic RM contribution for the entire line of sight through the Galaxy. 

\section{Depolarization canals in $P(\mathcal{R})$\ maps}\label{ch3_Sect: canals}
Dark, narrow channels (`canals') can be clearly seen in the images of Fig. \ref{ch3_stamps.fig}. Canals have been known to exist in $P(\lambda)$ images (e.g. Haverkorn et al. \cite{haverkorn00}), but, as far as we know, canals in $P(\mathcal{R})$ maps have only been mentioned by De Bruyn et al. (\cite{debruyn06}).

\begin{figure*}[!t]
\centering
\resizebox{\hsize}{!}{\includegraphics[width=8cm]{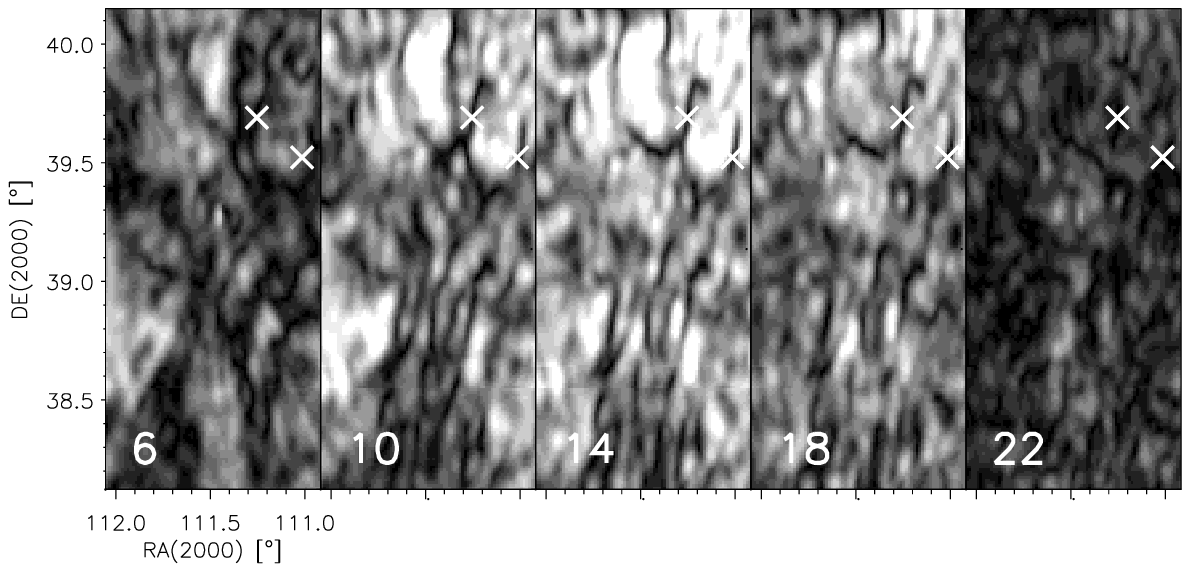}}
\resizebox{\hsize}{!}{\includegraphics[width=8cm]{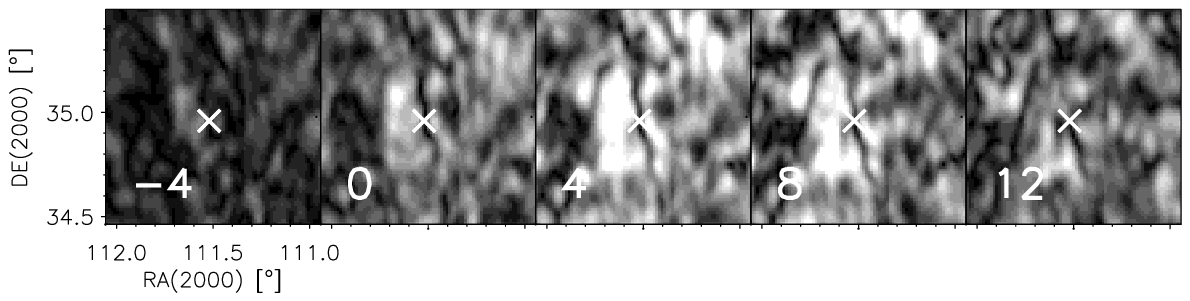}}
\caption{ Depolarization canals in the $P(\mathcal{R})$\ datacube. The $\mathcal{R}$\
  [rad/m$^2$] of each of the $P(\mathcal{R})$\ slices is indicated in the lower
  left. All maps have the same linear greyscale that saturates at 4.8
  K.  The area shown in the top row of this figure lies in the upper
  left of Fig. \ref{ch3_stamps.fig}. For the 2 pixels that are indicated by
  crosses, and that lie in depolarization canals, we plot the $P(\lambda^2)$\
  and $\Phi(\lambda^2)$\ spectra in the first row of Fig. \ref{ch3_channel_spectra.fig}
  (left cross, `CH 1' in Fig. \ref{ch3_channel_spectra.fig}) and middle row of
  Fig. \ref{ch3_channel_spectra.fig} (right cross, `CH 2'). The bottom row
  shows a canal in a different part of the mosaic, and the $P(\lambda^2)$\ and
  $\Phi(\lambda^2)$\ spectra of the canal pixel that is indicated by a cross are
  shown in the bottom row of Fig. \ref{ch3_channel_spectra.fig} (`CH 3').  
}
\label{ch3_r_channels.fig}
\end{figure*}

\begin{figure*}[!t]
\centering
\resizebox{\hsize}{!}{\includegraphics[width=8cm]{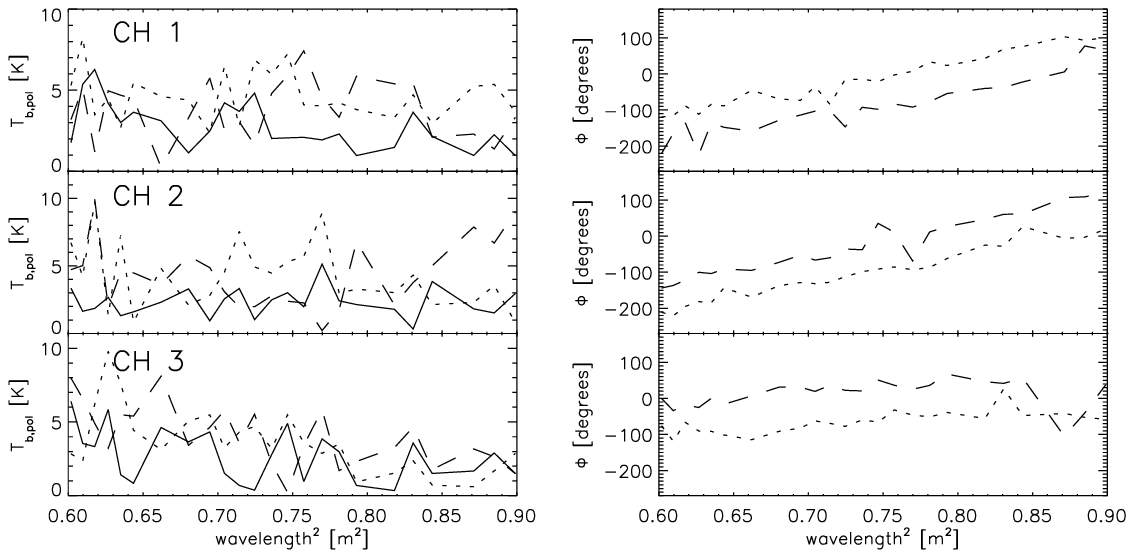}}
\caption{ Polarized brightness temperature T$_{\mathrm{b,pol}}$ [K]
  and polarization angle $\Phi$ [degrees] spectra for 3 canal pixels,
  and two adjacent pixels half a beamwidth from each canal pixel. Top
  row: canal pixel at ($\alpha,\delta$)=(111.24\degr,39.72\degr), and
  two adjacent lines of sight at $\alpha$=111.27\degr\ (dotted line),
  and $\alpha$=111.21\degr\ (dashed line). Middle row: canal pixel at
  ($\alpha,\delta$)=(110.98\degr,39.57\degr), and adjacent
  lines of sight at $\alpha$=111.01\degr\, dotted line, and
  $\alpha$=110.95\degr, dashed line. Bottom row: canal pixel at
  ($\alpha,\delta$)=(111.50\degr,34.94\degr), and two adjacent
  lines of sight at $\alpha$=111.53\degr, dotted line, and $\alpha$=
  111.47\degr, dashed line.  The 1$\sigma$ errors for polarized
  intensity and polarization angle are 2 K and 15\degr\ (for a S/N = 2
  detection) respectively Every 8$^{\mathrm{th}}$ channel is plotted. Bad
  frequency channels were selected using the same selection criteria
  as in Sect. \ref{ch3_Sect: data_characteristics}, and discarded.  }
\label{ch3_channel_spectra.fig}
\end{figure*}

In Fig. \ref{ch3_r_channels.fig} we show slices through the $P(\mathcal{R})$\ cube that have particularly strong depolarization canals. It appears from this figure that the canals essentially are not changing position with changing $\mathcal{R}$. This is demonstrated by the reference crosses that we positioned on three canals in this figure. In the first and second rows of Fig. \ref{ch3_channel_spectra.fig} we plot the $P(\lambda^2)$ (left column) and $\Phi(\lambda^2)$\ (right column) spectra for the two canal pixels from the top row of Fig.\ref{ch3_r_channels.fig} (`CH 1' and `CH 2' respectively). In the bottom row of Fig. \ref{ch3_channel_spectra.fig} we plot the spectra for the canal pixel in the bottom row of Fig. \ref{ch3_r_channels.fig}.

In Fig. \ref{ch3_channel_spectra.fig} we also plot the $P(\lambda^2)$\ and $\Phi(\lambda^2)$\ spectra for 2 lines of sight that lie on either side of the central canal pixel, at half a (synthesized) beamwidth away. These adjacent lines of sight share pixels with the line of sight that is centred on the canal pixel, and with these lines of sight we can investigate the nature of the canal pixels.
 
The $\Phi(\lambda^2)$\ spectra of the neighbouring lines of sight were unwrapped by fitting a straight line to the polarization angles by hand. As a first guess for the slope of this line, we used the $\mathcal{R}$ at which the $P(\mathcal{R})$ of the neighbouring lines of sight is high. Furthermore we required that for each line of sight the difference in polarization angle between the successive (plotted) wavelength$^2$ channels must be smaller than 90\degr. 

The difference in $\Phi(\lambda^2)$ for lines of sight adjacent to the canal pixels is large, often even almost 90\degr, averaged over the available $\lambda^2$. Also, the polarized brightness temperature of the 2 neighbouring lines of sight of each canal pixel are on average comparable. These 2 facts combined indicates that beam depolarization, i.e. summing polarization vectors with similar lengths but different orientations over the (synthesized) telescope beam, is an important factor in creating canals in the $P(\mathcal{R})$\ maps. 
In Haverkorn et al. (\cite{haverkorn00}) the same mechanism was identified to be producing canals in $P(\lambda^2)$\ maps. Line-of-sight depolarization, occurring for example as the nulls in Eqn. \ref{ch3_burn.eqn}, produces canals only at certain wavelength$^2$. Shukurov \& Berkhuijsen (\cite{shukurov03}) suggested that this mechanism is producing canals that are seen towards M31. This type of canal is characterised as extended along the $\mathcal{R}$ axis of the $P(\mathcal{R})$\ cube, contrary to the properties of the canal pixels we present here.


\section{The area around ($l,b$) = (109\degr$,$34.5\degr)}\label{ch3_Sect: blob}
The 2\degr$\times$2\degr\ area centred on ($l,b$) = (109.5\degr, 34.5\degr) is bright in polarized intensity (Fig. \ref{ch3_p_cr_derot_lm.fig}), shows a very uniform distribution in $\mathcal{R}$\ (Fig. \ref{ch3_cr_lm.fig}), and has a relatively small surface density of strong depolarization canals. Furthermore the $\chi^2_{\mathrm{red}}$\ fitted to the $\Phi(\lambda^2)$\ spectra are low (Fig. \ref{ch3_fscreens.fig}):
30\% of the lines of sight in this area have a $\chi^2_{\mathrm{red}}$ $<$ 2, and nearly 75\% have a $\chi^2_{\mathrm{red}}$ $<$ 3. There is no enhanced H$\alpha$ emission in this region (see the top panel of Fig. \ref{ch3_wham.fig}), and $B_{\|}$ varies only little (bottom panel of Fig. \ref{ch3_wham.fig}). The single-dish 408 MHz total intensity map by Haslam et al. (\cite{haslam82}) shows hardly an increase in brightness temperature in this region. There are not enough lines of sight in the Brouw \& Spoelstra (\cite{brouwspoelstra76}) data set to discern the area around ($l,b$) = (109.5\degr, 34.5\degr) from its surroundings.

The increased polarized brightness temperature with respect to its surroundings could be due to enhanced synchrotron emission, or a local decrease in depolarization, or to a combination of these. The polarized brightness temperature of this region is more than 3 K larger than that of its surroundings, and this would translate into a brightness temperature difference of at least 3 K/0.7 = 4.3 K, assuming that the radiation is 70\% polarized. Lower polarization percentages increase this brightness temperature contrast even further. An enhancement in synchrotron emission can be ruled out as an explanation for the increased $T_{\mathrm{b,pol}}$ of this area, since this is not seen in the Haslam data, and since beam averaging is not strong enough to attenuate the total intensity signal (the Haslam beam measures 0.85\degr$\times$0.85\degr).

The small variation in $\mathcal{R}$\ over this region, combined with an underabundance of strong depolarization canals, furthermore indicates that depolarization across the synthesized telescope beam is not a strong effect. This leaves a decrease in the amount of depolarization along the line of sight as an explanation for the observed increase in $T_{\mathrm{b,pol}}$. The low $\chi^2_{\mathrm{red}}$ of lines of sight in this area, in combination with secondary peaks that are not stronger than half the strength of the main peak, also indicates that the lines of sight in this area are not very complex, and line-of-sight depolarization effects are therefore not very important. We thus think that the origin of this particular region lies in a decreased amount of depolarization compared to its surroundings.

\section{Conclusions}
We applied Faraday tomography to high spectral-resolution radio polarization data to study the properties of the magnetised Galactic ISM. We showed that differential Faraday modulation in the foreground can shift the emitted angular frequency spectrum as a whole. In particular the 0-angular frequency from the emitted radiation is modulated in this way towards smaller angular scales, that are observable with an interferometer. We quantified how strong a linear gradient in $\mathcal{R}$ should be to accomplish this, and we showed that the gradients in $\mathcal{R}$\ that we find for the main peak in the $P(\mathcal{R})$\ spectra are indeed strong enough so that we are sensitive to the 0-angular frequency that is emitted at these Faraday depths. 

The main peak in the $P(\mathcal{R})$\ spectrum is not measurably resolved for 8\% of the lines of sight in our data set. An unresolved peak means that there is no synchrotron emission occurring in the Faraday-rotating region. This is unexpected, since cosmic rays that produce synchrotron emission pervade the ISM. We propose that the magnetic field orientation plays an important role in this, in that a magnetic field that is oriented along the line of sight produces only synchrotron emission perpendicular to the line of sight, hiding it from our view. By using the observed emission measures from the WHAM survey as input, and by simulating the thermal electron contribution to $\mathcal{R}$\ , we could determine the magnetic field strength along the line of sight, and we mapped this quantity over the area covered by our GEMINI data. The polarized point sources we found in our data have RMs that are comparable to the $\mathcal{R}$\ we find for the diffuse emission. In our $P(\mathcal{R})$\ maps we found depolarization canals, narrow structures with only a small fraction of the polarized intensity of adjacent lines of sight. We established that for a number of deep canals the polarization angle changes in many frequency channels by a large amount over the canal, in some cases 90\degr. We therefore argue that depolarization over the synthesized telescope beam is producing (at least) these canals. Finally, we investigated the properties of a large, conspicuous area in our data, and we argue that a decrease in depolarization along the line of sight as compared to its surroundings is probably responsible for the observational features of this region.

\section*{Acknowledgements}
The Westerbork Synthesis Radio Telescope is operated by the
Netherlands Foundation for Research in Astronomy (NFRA) with financial
support from the Netherlands Organization for Scientific Research
(NWO). The Wisconsin H-Alpha Mapper is funded by the National Science
Foundation.

\end{document}